\documentclass[a4,aps,prb,amsmath,twocolumn,showpacs,superscriptaddress,floatfix]{revtex4}

\usepackage{graphicx}
\usepackage[normal]{caption2}

\begin{document}

\title{Phase transitions in Ni$_{2+x}$Mn$_{1-x}$Ga with a high Ni excess}

\author{V.~V.~Khovaylo}
\email[Electronic address:\;]{v-khovaylo@cplire.ru}
\affiliation{Institute of Radioengineering and Electronics of RAS,
Moscow 125009, Russia}

\author{V.~D.~Buchelnikov}
\email[Electronic address:\;]{buche@csu.ru}
\affiliation{Physics
Faculty, Chelyabinsk State University, Chelyabinsk 454021, Russia}

\author{R.~Kainuma}
\affiliation{Department of Materials Science, Graduate School of
Engineering, Tohoku University, Sendai 980-8579, Japan}

\author{V.~V.~Koledov}
\affiliation{Institute of Radioengineering and Electronics of RAS,
Moscow 125009, Russia}

\author{M.~Ohtsuka}
\affiliation{Institute of Multidisciplinary Research for Advanced
Materials, Tohoku University, Sendai 980-8577, Japan}

\author{V.~G.~Shavrov}
\affiliation{Institute of Radioengineering and Electronics of RAS,
Moscow 125009, Russia}

\author{T.~Takagi}
\affiliation{Institute of Fluid Science, Tohoku University, Sendai
980--8577, Japan}

\author{S.~V.~Taskaev}
\affiliation{Physics Faculty, Chelyabinsk State University,
Chelyabinsk 454021, Russia}

\author{A.~N.~Vasiliev}
\affiliation{Low Temperature Physics Department, Moscow State
University, Moscow 119899, Russia}

\begin{abstract}
Ferromagnetic shape memory alloys Ni$_{2+x}$Mn$_{1-x}$Ga were
studied in the range of compositions $0.16 \le x \le 0.36$. The
experimental phase diagram, constructed from differential scanning
calorimetry, transport and magnetic measurements, exhibits
distinctive feature in a compositional interval $0.18 \le x \le
0.27$, where martensitic and magnetic transitions merge in a
first-order magnetostructural phase transition ferromagnetic
martensite $\leftrightarrow$ paramagnetic austenite. Observed in
this interval of compositions a nonmonotonous behavior of the
magnetostructural phase transition temperature was ascribed to the
difference in the exchange interactions of martensitic and
austenitic phase and to the competition between increasing number
of valence electron and progressive dilution of the magnetic
subsystem which occur in the presence of a strong magnetoelastic
interaction. Based on the experimental phase diagram, the
difference between Curie temperature of martensite $T_C^M$ and
Curie temperature of austenite $T_C^A$ was estimated. Influence of
volume magnetostriction was considered in theoretical modeling in
order to account for the existence of the magnetostructural phase
transition over a wide range of compositions.
\end{abstract}


\date{\today}

\maketitle

\section{Introduction}

In ferromagnetic shape memory alloys, structural (martensitic)
transition from high-temperature austenitic phase to
low-temperature martensitic phase takes place in ferromagnetically
ordered state. Combination of ferromagnetic properties of the
martensite and thermoelastic nature of the martensitic
transformation allows realization of a principle for operation of
shape and dimension in these materials. This can be achieved
either through switching of martensitic domains~\cite{1-u,2-u} or
through the shift of martensitic transition temperature~\cite{3-c}
by a magnetic field.

Among a variety of ferromagnetic shape memory
alloys~\cite{4-v,5-v} (see also Refs.~\onlinecite{6-o,7-s}), the
largest magnetic field-induced strain has been observed in
off-stoichiometric Ni-Mn-Ga single crystals.~\cite{8-s}
Observation of giant deformations induced by a magnetic field has
stimulated intensive studies of magnetic and structural properties
of Ni-Mn-Ga alloys. The results of these studies revealed a rich
phase diagram of this Heusler system. In particular,
stoichiometric or near-stoichiometric alloys undergo a first-order
premartensitic phase transition, resulting in a modulation of the
parent cubic structure.~\cite{9-z} Besides, phase transitions
between different crystallographic modifications of martensite can
be induced in off-stoichiometric Ni-Mn-Ga alloys by a change of
composition, temperature, or stress, or by the combination of
these parameters.~\cite{10-v}

Recent experimental studies revealed that along with the
phenomenon of large magnetic-field-induced strains Ni-Mn-Ga alloys
exhibit other properties of technological interest, specifically
the large magnetocaloric effect.~\cite{11-h,12-h,13-m,14-t,15-m}
Magnetic entropy change comparable with that recorded in so-called
giant magnetocaloric materials was observed at ambient
temperatures in Ni-Mn-Ga characterized by a coupled
magnetostructural phase
transition.~\cite{16-p,17-a,18-a,19-z,19a-z} It is notable that
the giant magnetocaloric materials are also characterized by the
simultaneously occurring magnetic and structural phase
transitions.~\cite{20-p,21-w,22-t}

Coupling of martensitic and magnetic transition temperatures,
$T_m$ and $T_C$, takes place in other ferromagnetic shape memory
alloy systems, such as Co-Ni-X (X = Al, Ga)~\cite{22a-o,22b-o} and
Ni-Fe-Ga.~\cite{6-o,22c-o} In Ni-Mn-Ga, this coupling seems to be
common and occurs for different cross-sections of the ternary
diagram. In Ni$_{2+x}$Mn$_{1-x}$Ga system, merging of $T_m$ and
$T_C$ was found to occur in a Ni$_{2.18}$Mn$_{0.82}$Ga
composition.~\cite{23-v} This effect has also been observed in the
alloys with substitution of Mn for Ga,~\cite{24-j}
Ni$_2$Mn$_{1+x}$Ga$_{1-x}$, and in the alloys where Ni atoms were
partially substituted for Ga.~\cite{25-l} Studies of
Ni$_{2+x}$Mn$_{1-x}$Ga and Ni$_2$Mn$_{1+x}$Ga$_{1-x}$ revealed
similar tendency of $T_m$ to increase and $T_C$ to decrease with
the deviation from stoichiometry. Increase of $T_m$ in these alloy
systems is attributable to the increase in electron concentration
$e/a$, i.e., to the Hume-Rothery mechanism. Although first
principles calculation of non-stoichiometric Ni$_2$MnGa alloys
indicated~\cite{26-m} that substitution-induced change in
electronic structure did not fall into a rigid band filling
scenario, empirical dependence between electron concentration and
martensitic transition temperature was found to hold for a large
number of compositions,~\cite{27-c} suggesting good applicability
of the rigid band model. Decrease of $T_C$ observed in
Ni$_{2+x}$Mn$_{1-x}$Ga and Ni$_2$Mn$_{1+x}$Ga$_{1-x}$ has
presumably different origin. Since in Ni-Mn-Ga alloys magnetic
moment of $\sim 4~\mu_B$ is located on Mn atoms, lowering of $T_C$
in the Ni$_{2+x}$Mn$_{1-x}$Ga system can reasonably be explained
as caused by the dilution of the magnetic subsystem. For the
system with Mn excess, Ni$_2$Mn$_{1+x}$Ga$_{1-x}$, weakening of
exchange interaction could be accounted for by antiferromagnetic
coupling of excessive Mn atoms,~\cite{28-e} although such a
picture has to be verified experimentally yet. Systematic study of
magnetic properties of Ni$_{2+x}$Mn$_{1-x}$Ga alloys~\cite{29-k}
showed that both the interatomic distances and the overlap of
electronic orbitals play an important role in the change of
exchange parameters at structural transition, and that the
exchange interactions are stronger in the martensitic state.

The difference in Curie temperature of martensite and
austenite~\cite{29-k} leads to unusual magnetic properties of
compounds with merged magnetic and structural transition
temperature. Higher Curie temperature of martensite as compared to
that of austenite and intrinsic thermal hysteresis of the
martensitic transition results in a well-defined temperature
hysteresis seen on temperature dependencies of magnetization
$M(T)$ in these alloys.~\cite{16-p,18-a,30-k} Isothermal
magnetization measurements of Ni$_{2.19}$Mn$_{0.81}$Ga revealed
marked metamagnetic-like anomalies on field dependencies of the
magnetization.~\cite{31-f} They can be either reversible or
irreversible, depending on the temperature of the measurements,
and are caused by the field-induced transitions from paramagnetic
austenite to ferromagnetic martensite.

\begin{figure}[t]
\includegraphics[width=\columnwidth]{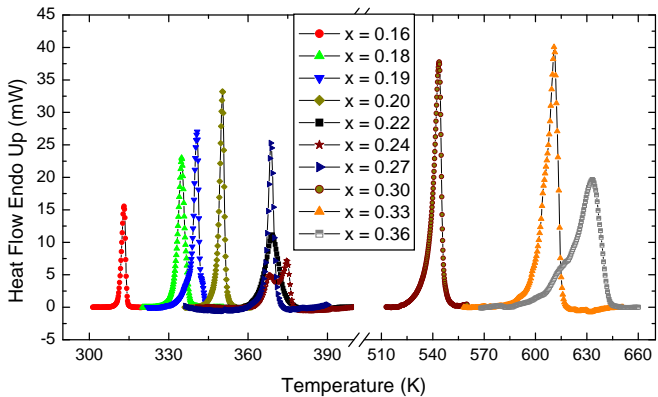}
\caption{Heating DSC scans showing latent heat of transition from
martensite to austenite in Ni$_{2+x}$Mn$_{1-x}$Ga $(0.16 \le x \le
0.36)$.}
\end{figure}

So far, phase transitions in the Ni$_{2+x}$Mn$_{1-x}$Ga system
were studied in the range of compositions $0 \le x \le
0.20$.~\cite{16-p,17-a,18-a,23-v,29-k,30-k,31-f,31a-m,31b-k}
Studies of Ni$_{2+x}$Mn$_{1-x}$Ga alloys with a higher Ni excess
are motivated by several reasons. First, it is likely that $T_m$
and $T_C$ are still merged in the alloys with $x \ge 0.20$. For
better understanding of the phenomenon of coupled
magnetostructural transition it is necessary to determine the
complete composition interval where it is observed. Since such
alloys show attractive magnetocaloric
properties,~\cite{16-p,17-a,18-a,19-z,19a-z} this information can
also be of technological interest. Second, the phenomenological
theory of phase transitions in Ni$_{2+x}$Mn$_{1-x}$Ga
predicts~\cite{23-v,32-b} that the magnetostructural transition is
realized in $0.18 \le x \le 0.20$ interval; in the alloys with a
higher Ni excess $T_m$ becomes higher than $T_C$ and the
martensitic transformation takes place in paramagnetic state.
Construction of the experimental phase diagram for the alloys with
high Ni excess provides good opportunity for verification of this
theoretical prediction. For this aim we studied
Ni$_{2+x}$Mn$_{1-x}$Ga alloys characterized by the Ni excess $0.16
\le x \le 0.39$.

\section{Sample preparation and measurements}

Polycrystalline ingots with nominal compositions in the mentioned
above range of $x$ were prepared by a conventional arc-melting
method. Since the weight loss during arc-melting was small
$(<0.2$\%) we assume that the real compositions correspond to the
nominal ones. The ingots were annealed at 1100~K for 9~days and
quenched in ice water. Metallographic studies revealed a single
phase state in all the compositions except the alloy with the
highest Ni excess, $x = 0.39$. For this composition, optical
observation showed the presence of a secondary phase indicating
that a phase segregation takes place for alloys with $x \ge 0.39$.
Because of this, discussion of the experimental results obtained
will be restricted for alloys with $x \le 0.36$.

Samples for calorimetric, transport and magnetic measurements were
cut from the middle part of the ingots. Characteristic
temperatures of the direct and reverse martensitic transformations
were determined from differential scanning calorimetry (DSC)
measurements, performed with a heating/cooling rate 5~K/min. Curie
temperature $T_C$ was determined from temperature dependencies of
magnetization, $M(T)$, measured by a vibrating sample magnetometer
in a magnetic field $H = 0.01$~T with a heating/cooling rate
2~K/min. For the compositions which exhibit first order magnetic
phase transition, $T_C$ was determined as an average between
values observed at heating and cooling. Isothermal magnetization
was measured at liquid helium temperature in magnetic fields up to
5~T by a Quantum Design superconducting quantum interference
device (SQUID) magnetometer. Magnetization saturation was
determined by a linear extrapolation of $M(H)$ dependencies from
the high fields. Measurements of the thermal expansion coefficient
were performed by a strain gage technique. A strain gage was glued
to the carefully polished flat surface of sample. The measurements
were done in a temperature interval from 300 to 380~K with a
heating rate 1~K/min.

\section{Experimental results}

DSC measurements provide simple and effective tool to detect
martensitic transformations. Well-defined peaks seen on cooling
and heating DSC curves correspond to direct and reverse
martensitic transformation, respectively. Direct martensitic
transformation is characterized by martensite start ($M_s$) and
martensite finish ($M_f$) temperatures. Accordingly, reverse
martensitic transformation can be characterized by austenite start
($A_s$) and austenite finish ($A_f$) temperatures. As an example
of these measurements, DSC heating scans for the alloys from
studied compositional interval are shown in Fig.~1. Complex
transformation behavior observed in $x = 0.33$, $x = 0.36$ and,
especially, in $x = 0.24$ samples is presumably caused by the
coexisting martensitic phases which transform to the austenitic
state at slightly different temperatures. Describing experimental
results, we will use the thermodynamic equilibrium temperature of
martensitic transformation, $T_m$, determined as $T_m = (M_s +
A_f)/2$ (Ref.~\onlinecite{33-t}).

\begin{figure}[t]
\includegraphics[width=\columnwidth]{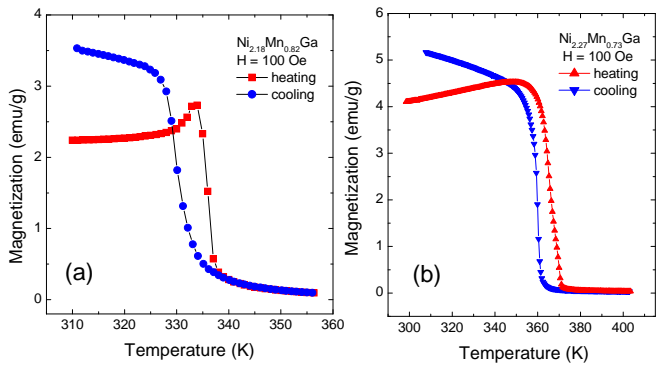}
\caption{Temperature hysteresis of the ferromagnetic transition
observed in Ni$_{2+x}$Mn$_{1-x}$Ga $(0.18 \le x \le 0.27)$ alloys
undergoing coupled magnetostructural phase transition.}
\end{figure}

Our DSC measurements revealed general tendency of the martensitic
transformation temperature to increase with Ni excess $x$ which is
caused by the increase in the number of valence electrons. The
increase of $T_m$ is, however, non-monotonous. In a compositional
interval $0.16 \le x \le 0.22$ the martensitic transformation
temperature increases from $\sim 310$~K $(x = 0.16)$ to $\sim
370$~K $(x = 0.22)$. Further modification of $x$ from $x = 0.22$
to $x = 0.27$ has no essential effect on $T_m$ which remains
essentially constant, $T_m \approx 370$~K, in this compositional
interval. A jump-like increase of the transformation temperature
is observed as the composition changes from $x = 0.27$ to $x =
0.30$. In the compositions with $x \ge 0.33$ the martensitic
transformation takes place at temperatures above 600~K (Fig.~1).

Since previous studies of Ni$_{2+x}$Mn$_{1-x}$Ga $(0 \le x \le
0.20)$~\cite{23-v,29-k} showed that the martensitic and
ferromagnetic transition temperatures merge in the $x = 0.18$
composition, the observed non-monotonous behavior of $T_m$
(Fig.~1) could be related to the coupling of martensitic and
ferromagnetic transitions. For the alloys studied, Curie
temperature $T_C$ was determined as a minimum on the temperature
derivative of the magnetization curve, $dM/dT$, measured in a
field of 0.01~T. Results of these measurements revealed that $T_C
\approx T_m$ in the interval of compositions $0.18 \le x \le
0.27$. Moreover, $T_C$ in these alloys exhibits pronounced
hysteretic feature as is shown in Fig.~2 for the case of end
members of this compositional interval. Because of the temperature
hysteresis, $T_C$ of these compounds was determined as the average
between values measured at heating and cooling.

\begin{figure}[b]
\includegraphics[width=\columnwidth]{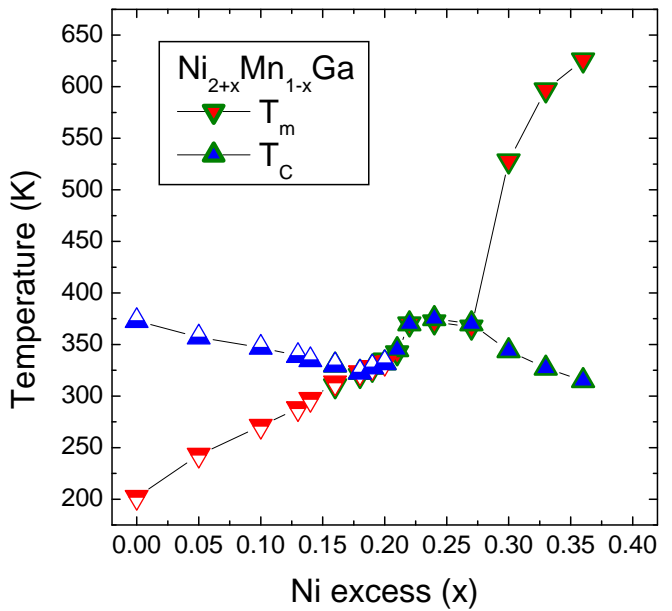}
\caption{Experimental phase diagram of Ni$_{2+x}$Mn$_{1-x}$Ga $(0
\le x \le 0.36)$ constructed from DSC and magnetization
measurements. Half-filled triangles are results from
Ref.~\onlinecite{23-v}.}
\end{figure}

The phase diagram of Ni$_{2+x}$Mn$_{1-x}$Ga in the studied
compositional interval, constructed from the DSC and magnetization
measurements, is shown in Fig.~3. Three different regions can be
distinguished on this phase diagram. The first region is
characterized by the Ni excess $x \le 0.16$. In this region $T_C >
T_m$ and the martensitic transformation takes place when in the
ferromagnetic state. Alloys from the second region with the Ni
excess $0.18 \le x \le 0.27$ are characterized by a coupled
magnetostructural transition, i.e., $T_m \approx T_C$.
Ferromagnetic transition in this compositional interval has a
characteristic of a first-order phase transition, showing
pronounced hysteresis on temperature and field dependencies of
magnetization, $M(T)$ and $M(H)$.~\cite{30-k,31-f} Such unusual
magnetic properties of these alloys have been attributed to
simultaneously occurring martensitic and ferromagnetic
transitions.~\cite{4-v,30-k} Finally, the third region is
characterized by a high martensitic transformation temperature,
$T_m > 550$~K, and a low Curie temperature, $T_C < 350$~K. In this
region, with the Ni excess $x \ge 0.30$, the martensitic
transformation takes place when in the paramagnetic state. The
occurrence of martensitic transformation at high temperatures
makes alloys from this region attractive for application as
high-temperature shape memory alloys.

Since substitution of Ni for Mn results in the dilution of the
magnetic subsystem, the observed increase of $T_C$ in the $0.18
\le x \le 0.22$ alloys manifests a strong interrelation between
magnetic and structural subsystems in Ni$_{2+x}$Mn$_{1-x}$Ga. In
order to check whether magnetic moment has an anomalous
compositional dependence in this range of $x$, field dependencies
of magnetization $M(H)$ were measured at 5~K. Along with
Ni$_{2+x}$Mn$_{1-x}$Ga $(0.16 \le x \le 0.36)$, $M(H)$ was also
measured on used in the previous studies~\cite{23-v,29-k} samples
with a smaller deviation from stoichiometry and for the
stoichiometric Ni$_2$MnGa. Calculated from these measurements
magnetic moment per formula unit as a function of Ni excess $x$ is
shown in Fig.~4.  The results obtained indicate that within the
experimental error the magnetic moment decreases approximately
linearly upon substitution of Mn for Ni in the interval of
compositions $0 \le x \le 0.36$.

\begin{figure}[t]
\includegraphics[width=\columnwidth]{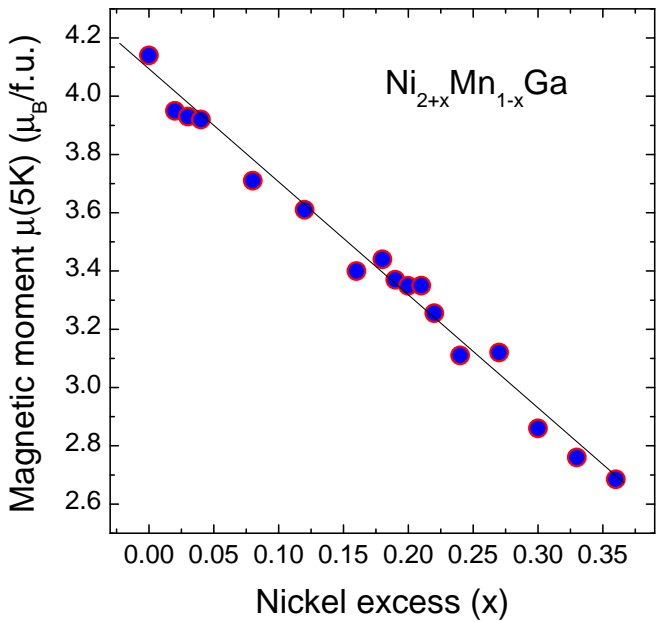}
\caption{Magnetic moment $\mu_B$ per formula unit as a function of
Ni excess $x$ in Ni$_{2+x}$Mn$_{1-x}$Ga.}
\end{figure}

\begin{figure}[t]
\includegraphics[width=\columnwidth]{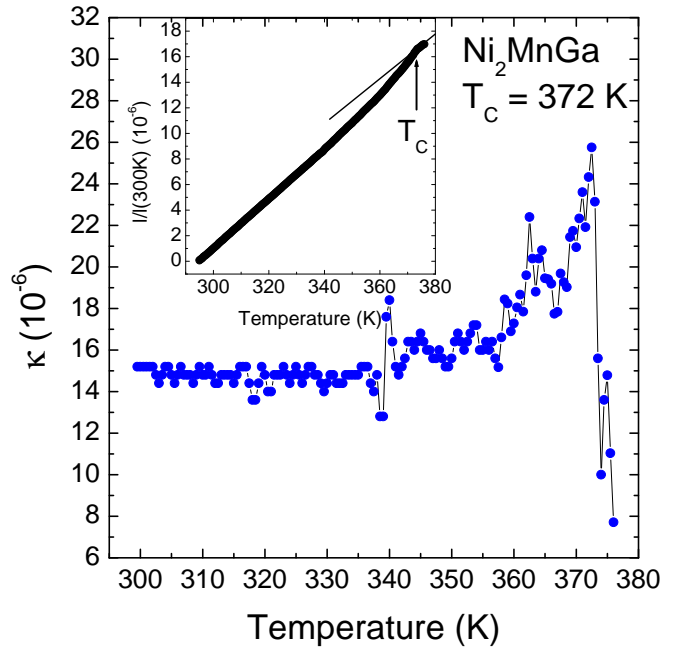}
\caption{Temperature dependence of the thermal expansion
coefficient in cubic austenitic phase of stoichiometric
Ni$_2$MnGa. Shown in the inset is a thermal expansion curve of
this alloy.}
\end{figure}

The result on the thermal expansion measurement is shown in the
inset of Fig.~5. As the temperature is lowered through the Curie
temperature, a small drop in the volume $\omega = 3(\Delta l/l)$
of the specimen is observed indicating that the exchange striction
in Ni$_2$MnGa is negative. Similar behavior has been found in
other Mn-containing Heusler alloys.~\cite{33a-c} Temperature
dependence of the thermal expansion coefficient $\kappa$,
determined from this measurement, is shown in Fig.~5. A noticeable
increase in $\kappa$ is observed at temperatures above ~ 340 K.
$\kappa$ exhibits a jump-like drop at Curie temperature $T_C =
372$~K, when the sample transforms to the paramagnetic state.

\section{Influence of volume magnetostriction on the phase
diagram of Ni$_{2+x}$Mn$_{1-x}$Ga alloys}

The influence of anisotropic magnetostriction on structural phase
transitions in cubic ferromagnets was theoretically studied in
several works (see Ref.~\onlinecite{4-v}, and references therein).
Taking into consideration the magnetoelastic interactions, it was
found that a coupled magnetic and structural (magnetostructural)
phase transition exists on the phase diagram of
Ni$_{2+x}$Mn$_{1-x}$Ga alloys. Theoretical calculations of the $T
- x$ phase diagram revealed, however, that the magnetostructural
transition is realized in rather a narrow concentration interval.
Contrary to this, the experimental phase diagram reported in this
work evidences that $T_m$ and $T_C$ are merged in a considerable
interval of compositions, from $x = 0.18$ to $x = 0.27$ (Fig.~3).
Since such alloys undergo a transformation from paramagnetic
austenite to ferromagnetic martensite, it can be suggested that
the influence of the anisotropic magnetostriction can be neglected
in the composition interval where the magnetostructural phase
transition is realized. Instead, contribution from the volume
magnetostriction, which is usually large in the vicinity of
magnetic phase transitions, should be considered when describing
phase transitions in the framework of Landau's theory.

For description of a phase transition from cubic paramagnetic to
cubic ferromagnetic phase let us consider a Ginzburg-Landau
functional containing order parameters responsible for structural
and magnetic phase transitions and for the volume change at
temperatures close to Curie temperature $T \sim
T_C$~\cite{34-f,35-i}

\begin{widetext}

\begin{equation}
\begin{split}
F  &= -Ae_1 + \frac{1}{2}A_0e^2_1 + \frac{1}{2}a_1(e^2_2+e^2_3) +
De_1(e^2_2+e^2_3) + \frac{1}{3}be_3(e_3^2-3e_2^2)
+\frac{1}{4}c_1(e_2^2+e_3^2)^2 +\\
&+\frac{1}{\sqrt{3}}B_1e_1\mathbf{m}^2 +
+K(m_x^2m_y^2+m_y^2m_z^2+m_z^2m_x^2)+\frac{1}{2}\alpha_1\mathbf{m}^2
+ \frac{1}{4}\delta_1\mathbf{m}^4 + Pe_1.
\end{split}
\end{equation}

\end{widetext}

\noindent Here $e_i$ are the linear combinations of the
deformation tensor components,
$e_1=(e_{xx}+e_{yy}+e_{zz})/\sqrt{3}$,
$e_2=(e_{xx}-e_{yy})/\sqrt{2}$,
$e_3=(2e_{zz}-e_{yy}-e_{xx})/\sqrt{6}$; $A$ is a coefficient
proportional to the thermal expansion coefficient,
$A_0=(c_{11}+2c_{12})/\sqrt{3}$ is the bulk modulus, $a_1$, $b$,
$D$, $c_1$ are the linear combinations of second-, third-, and
fourth-order elastic moduli, respectively, $a_1 = c_{11}-c_{12}$,
$b=(c_{111}-3c_{112}+2c_{123})/6\sqrt{6}$,
$D=(c_{111}-c_{123})/2\sqrt{3}$,
$c_1=(c_{1111}+6c_{1112}-3c_{1122}-8c_{1123})/48$; $\textbf{m} =
\textbf{M}/M_0$ is the magnetization vector ($M_0$ is the
magnetization saturation), $B_1$ is the volume (exchange)
magnetostriction constant, $K$ is the first cubic anisotropy
constant, $\alpha_1$ and $\delta_1$ are the exchange constants,
and $P$ is the hydrostatic pressure.

Minimization of Eq.~(1) with respect to $e_1$ leads to the
following result:

\begin{equation}
e_1 =
\frac{A}{A_0}-\frac{D}{A_0}(e_2^2+e_3^2)-\frac{1}{\sqrt{3}}\frac{B_1}{A_0}\mathbf{m}^2
- \frac{P}{A_0} .
\end{equation}

\noindent After substitution of Eq.~(2) into Eq.~(1) the
expression for the free energy of ferromagnet is

\begin{widetext}

\begin{equation}
\begin{split}
F &= - \frac{(A - P)^2}{2A_0}+\frac{1}{2}a(e_2^2+e_3^2) +
\frac{1}{3}be_3(e_3^2-3e_2^2) + \frac{1}{4}c(e_2^2+e_3^2)^2 -\\
&-\frac{1}{2}B\mathbf{m}^2(e_2^2+e_3^2)
+K(m_x^2m_y^2+m_y^2m_z^2+m_z^2m_x^2) +
\frac{1}{2}\alpha\mathbf{m}^2 + \frac{1}{4}\delta\mathbf{m}^4,
\end{split}
\end{equation}

\noindent where

\begin{equation}
%\begin{split}
a=a_1+2\frac{(A-P)D}{A_0}, \;\; c=c_1-2\frac{D^2}{A_0}, \;\;
B=2\frac{DB_1}{\sqrt{3}A_0},  \;\;
\alpha=\alpha_1+2\frac{(A-P)B_1}{\sqrt{3}A_0}, \;\;
\delta=\delta_1-2\frac{B_1^2}{3A_0}.
%\end{split}
\end{equation}

\end{widetext}

As evident from Eq.~(4), coefficient $B$ is proportional to the
volume magnetostriction constant $B_1$. For the sake of
definiteness we assume in further discussion that $B > 0$, the
generalized third- and fourth-order elastic moduli $b, c > 0$, the
magnetic anisotropy constant $K > 0$, and the exchange constant
$\delta > 0$.

Minimization of Eq.~(3) with respect to $e_{2,3}$ and $m_{x,y,z}$
leads to the following equilibrium phases of the ferromagnet.

\noindent (I) Cubic paramagnetic phase (PC)

\begin{equation}
m_x = m_y = m_z = 0, \qquad e_2 = e_3 = 0
\end{equation}

\noindent is stable at $\alpha \ge 0$, $a \ge 0$.

\noindent (II) Tetragonal paramagnetic phase (PT)

\begin{equation}
m_x = m_y = m_z = 0, \quad e_2 = 0, \quad e_3 =
-\frac{b+\sqrt{b^2-4ac}}{2c}
\end{equation}

\noindent is stable at

\begin{equation}
\alpha \ge \frac{Bb^2}{4c^2}, \quad a \le \frac{b^2}{4c}, \quad a
\ge b\sqrt{\frac{\alpha}{B}}-\frac{c\alpha}{B}.
\end{equation}

\noindent (III) Cubic ferromagnetic phase (FC) with magnetization
vector along [001]

\begin{equation}
m_x = m_y = 0, \quad m_z^2 = -\frac{\alpha}{\delta}, \quad e_2 =
e_3 = 0
\end{equation}

\noindent is stable at

\begin{equation}
-\delta \le \alpha \le 0, \qquad  a \ge -\frac{B\alpha}{\delta}.
\end{equation}

\noindent (IV) Tetragonal collinear ferromagnetic phase (FT) with
magnetization vector along [001] axis

\begin{equation}
m_x = m_y = 0, \qquad m_z^2 = -\frac{\alpha-Be_3^2}{\delta}
\end{equation}

\noindent and deformations

\begin{equation}
\begin{split}
& e_2 = 0, \\ & e_3 =
-\frac{b+\sqrt{b^2-4(a+B\alpha/\delta)(c-B^2/\delta)}}{2(c-B^2/\delta)}
\end{split}
\end{equation}

\noindent is stable at

\begin{equation}
\begin{split}
 a \le &\frac{b^2}{4(c-B^2/\delta)}-B\frac{\alpha}{\delta},
\quad \alpha \ge -\delta+\frac{Bb^2}{4(c-B^2/\delta)^2}, \\ &a \le
b\sqrt{\frac{\alpha}{B}}-\frac{c\alpha}{B}, \qquad \alpha \ge
\frac{Bb^2}{4(c-B^2/\delta)^2}.
\end{split}
\end{equation}

It follows from the symmetry consideration that, beside these
states, others equilibrium phases having the same energy and areas
of stability can be realized in the ferromagnet. These are
tetragonal paramagnetic phases with deformations along [100] and
[010] axes, ferromagnetic cubic phases with magnetization vectors
along [100] and [010] axes, and tetragonal phases with
deformations and magnetization vectors along [100] and [010] axes.

The lines of the phase transitions between states (I)-(IV) can be
found from the conditions of the phase equilibria. They are
determined by the following expressions

\begin{equation}
\begin{split}
&\mathrm{I\!\leftrightarrow\!II}\!: \; a = \frac{2b^2}{9c}, \\
&\mathrm{I\!\leftrightarrow\!III}\!:\alpha=0,\\
&\mathrm{I\!\leftrightarrow\!IV}\!:\frac{1}{3}be_3^3+\big(\frac{\alpha
B}{\delta}+a\big)e_3^2-\frac{\alpha^2}{\delta}=0, \\ &\text{with}
\;  e_3 \; \text{from Eq.~(11)} \\
&\mathrm{III\!\leftrightarrow\!IV}\!: a =
\frac{2b^2}{9(c-B^2/\delta)}-\frac{B\alpha}{\delta}.
\end{split}
\end{equation}

\noindent On the $a-\alpha$ diagram, the coordinates of starting
and ending points of magnetostructural phase transition are

\begin{equation}
\text{S} \Big[\frac{2b^2}{9(c - B^2/\delta)}, 0\Big], \;\;
\text{E} \Big[\frac{2b^2}{9c}, \frac{4Bb^2}{9c^2}\Big].
\end{equation}

In order to compare the results of the calculations with the
experimental data, the  $a - \alpha$ phase diagram can be
represented in $T - x$ coordinates. Let us assume that $P=0$.
According to the Ginzburg-Landau theory, in the vicinity of Curie
point $T_C$ the exchange parameter $\alpha$ can be represented as

\begin{equation}
\alpha = \alpha_0\frac{T - T_C(x)}{T_{C0}}.
\end{equation}

\noindent In the vicinity of structural phase transition point
$T_m$ the generalized second-order elastic modulus $a$ can be
presented as

\begin{equation}
a = a_0\frac{T - T_m(x)}{T_{m0}}.
\end{equation}

In Eqs.~(15) and (16) we assume simple linear compositional
dependencies of $T_C$ and $T_m$ temperatures~\cite{23-v}

\begin{equation}
T_C(x) = T_{C0}-\gamma x, \quad T_m(x) = T_{m0} + \sigma x,
\end{equation}

\noindent where $T_{C0}$ and $T_{m0}$ are temperatures of
ferromagnetic and martensitic phase transitions for the
stoichiometric composition, $\gamma$ and $\sigma$ are coefficients
determined from the experiments. Substitution of Eqs.~(15)-(17)
into expression for the coordinates of the critical points S and E
(14) gives coordinates of those points on the $T - x$ phase
diagram

\begin{equation}
\begin{split}
& x_S = \frac{1}{\gamma+\sigma}\bigg(T_{C0}-T_{m0}-
\frac{2}{9}\frac{b^2T_{m0}}{a_0(c-B^2/\delta)}\bigg), \\ &
T_S=T_{C0}-\gamma x_S, \\ & x_E =
\frac{1}{\gamma+\sigma}\bigg(T_{C0}-T_{m0}^{\prime}+
\frac{4}{9}\frac{Bb^2T_{C0}}{\alpha_0c^2}\bigg),
\\ & T_E=T_{m0}^{\prime}+\sigma x_E, \;
 T_{m0}^{\prime} = T_{m0}\bigg(1+\frac{2}{9}\frac{b^2}{ca_0}\bigg).
\end{split}
\end{equation}

The lines of phase transitions (13) between states (I)-(IV) in the
$T - x$ coordinates will be as follows

\begin{equation}
\begin{split}
&\mathrm{I\!\leftrightarrow\!II}\!: \; T = -T_{m0}^{\prime} +
\sigma x \\ &\mathrm{I\!\leftrightarrow\!III}\!: \; T = T_{C0} -
\gamma x \\ &\mathrm{I\!\leftrightarrow\!IV}\!: \; T \approx
\tilde{T}_{m0} + \tilde{\sigma} x \\
&\mathrm{II\!\leftrightarrow\!IV}\!: \; T \approx \tilde{T}_{C0} -
\tilde{\gamma}x\\ &\mathrm{III\!\leftrightarrow\!IV}\!: \; T =
\tilde{T}_{m0} + \tilde{\sigma}x, \\ & \tilde{T}_{m0} =
T_{m0}\frac{(1+2b^2)/[9ca_0(1-B^2/c\delta)]+B^2\alpha_0/a_0\delta}{1+B\alpha_0T_{m0}/T_{C0}a_0\delta},\\
 & \tilde{\sigma}=\sigma\frac{1-B\alpha_0T_{m0}\gamma/\sigma T_{C0}a_0\delta}{1+B\alpha_0T_{m0}/T_{C0}a_0\delta}.
\end{split}
\end{equation}

The expressions for the lines of phase transitions
$\mathrm{I\!\leftrightarrow\!IV}$ and
$\mathrm{II\!\leftrightarrow\!IV}$ are written in the first linear
approximation relatively to temperature and composition. The
values of $\tilde{T}_{C0}$  and $\tilde{\gamma}$  were estimated
by numerical calculations from Eq.~13 (see below). Line
$\mathrm{I\!\leftrightarrow\!IV}$ is the line of coupled
magnetostructural phase transition with $T_m = T_C$.

To construct the $T - x$ phase diagram we need to determine the
value of magnetostriction constants $B_1$ and $B$. For this aim we
shall consider the phase transition line
$\mathrm{I\!\leftrightarrow\!III}$.

It follows from Eqs.~(5) and (8) that $m = 0$ in cubic
paramagnetic phase and $m^2 = - \alpha/\delta$ in cubic
ferromagnetic phase. In the vicinity of ferromagnetic phase
transition, constant $A$ can be written as
$A=\kappa_0A_0(T-T_C)$,~\cite{34-f} where $\kappa_0$ is the
thermal expansion coefficient in paramagnetic phase. In this case,
constant $\alpha_1$ from Eq.~(4) can be represented as

\begin{equation}
\alpha_1 = \alpha_{10}\frac{(T-T_C)}{T_C}
\end{equation}

\noindent and, therefore,

\begin{equation}
\begin{split}
& \alpha = \alpha_{0}\frac{(T-T_C^{\prime})}{T_C}, \\ & \alpha_0 =
\alpha_{10}+\frac{2}{\sqrt{3}}\kappa_0B_1T_C, \\ & T_C^{\prime} =
T_C\bigg(1+\frac{2}{\sqrt{3}}\frac{PB_1}{A_0\alpha_0}\bigg).
\end{split}
\end{equation}

Using expressions for $e_1$ and $m$ from Eqs.~(2), (5), and (8)
and assuming that $P \ne 0$ we can obtain from Eq.~(3) the thermal
expansion coefficient in paramagnetic and ferromagnetic phases,
$\kappa_p$ and $\kappa_f$:

\begin{equation}
\kappa_p = \frac{\partial^2F}{\partial P\partial T} = \kappa_0,
\end{equation}

\begin{equation}
\kappa_f = \kappa_0 +
\frac{1}{\sqrt{3}}\frac{B_1}{A_0T_C}\frac{\alpha_0}{\delta}.
\end{equation}

\noindent The jump of thermal expansion coefficient at Curie point
is

\begin{equation}
\Delta\kappa = \kappa_f - \kappa_p =
\frac{1}{\sqrt{3}}\frac{B_1}{A_0T_C}\frac{\alpha_0}{\delta}.
\end{equation}

\noindent Experimentally measured thermal expansion coefficient of
Ni$_2$MnGa (Fig.~5) indicates that $\Delta\kappa =
1.8\times10^{-5}$~K$^{-1}$ at $T=T_C=372$~K. Using reported in
Ref.~\onlinecite{36-m} experimental values of elastic moduli
$c_{11} = 136$~GPa, $c_{12} = 92$~GPa and considering that at $T
\ll T_C$ and $P = 0$ magnetization module $m \approx 1$ (i.e.,
$\alpha_0/\delta \approx 1$), the volume magnetostriction is

\begin{equation}
B_1 \approx \sqrt{3}\Delta\kappa A_0T_C \approx 21\times10^9
\text{erg/cm}^3.
\end{equation}

\noindent Considering the definition of the $B$ parameter
[(Eq.~4)] and assuming that the ratio $D/A_0$ in Ni$_2$MnGa is
similar to that in Cu-based shape memory
alloys,~\cite{36a-g,36b-g} $D/A_0 \approx 10$, we obtain that $B
\approx 2.4\times10^{11}$~erg/cm$^3$.

To construct the $T - x$ phase diagram we use the following values
for the remaining parameters: $\alpha_0 = 5\times
10^9$~erg/cm$^3$, $\delta = 5\times 10^9$~erg/cm$^3$,
$a_0=5\times10^{11}$~erg/cm$^3$, $b=5\times10^{12}$~erg/cm$^3$,
$c=5\times10^{13}$~erg/cm$^3$; for $0\le x \le 0.18$: $\sigma =
860$~K, $\gamma = 280$~K, $T_{C0}=372$~K, $T_{m0}=132$~K; for
$0.27 \le x \le 0.36$: $\sigma = 3400$~K, $\gamma = 2600$~K,
$T_{C0} = 1050$~K, and $T_{m0} = 450$~K. For the given values of
the parameters, empirical expressions for the lines of phase
transitions are

\begin{equation}
\begin{split}
\nonumber
 &\mathrm{I\!\leftrightarrow\!II}\!: \; T = -550 + 3400x,
0.27 \le x \le 0.36\\ &\mathrm{I\!\leftrightarrow\!III}\!: T = 372
- 280x, 0 \le x \le 0.18 \\ &\mathrm{I\!\leftrightarrow\!IV}\!: T
\approx 200 + 694x, 0.18 \le x \le 0.27 \\
&\mathrm{II\!\leftrightarrow\!IV}\!: T \approx 538 - 626x, 0.27
\le x \le 0.36 \\ &\mathrm{III\!\leftrightarrow\!IV}\!: T = 200 +
694x, 0 \le x \le 0.18.
\end{split}
\end{equation}

Theoretically constructed $T - x$ phase diagram is shown in
Fig.~6. Note that a strong deviation from linear dependencies of
the phase transition line $\mathrm{I\!\leftrightarrow\!IV}$ is
observed on the experimental phase diagram for the compositions
near $0.22 \le x \le 0.27$. Because of this, the line of this
transition is shown in Fig.~6 schematically. The theoretically
constructed $T - x$ phase diagram presented in Fig.~6 is in a
qualitative agreement with the experimental one (Fig.~3).

\begin{figure}[t]
\includegraphics[width=6cm]{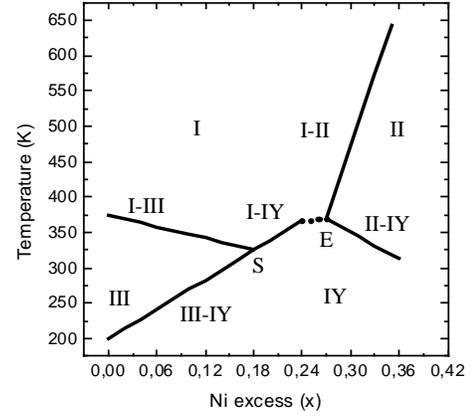}
\caption{Theoretical phase diagram of a cubic ferromagnet in $T-x$
coordinates.}
\end{figure}

For the indicated values of the parameters and Eq.~(18) follows
that the compositional interval of the magnetostructural phase
transition is

\begin{equation}
\begin{split}
\nonumber
 x_E - x_S & = \frac{4Bb^2}{9c^2\alpha_0(\sigma+\gamma)}\bigg[T_{C0} +
 T_{m0}\frac{B\alpha_0}{2a_0\delta(1-B^2/c\delta)}\bigg] \approx \\
& \approx 3.22 \times 10^{-12}B = 0.08.
\end{split}
\end{equation}

\section{Discussion}

The most interesting feature of the experimental phase diagram is
the coupling of martensitic and ferromagnetic transitions in
rather a wide composition interval, from $x = 0.18$ to $x = 0.27$.
Moreover, the coupled magnetostructural phase transition exhibits
a non-monotonous dependence on the Ni excess $x$ (Fig.~3).
Decoupling of $T_m$ and $T_C$ in $x = 0.30$ is accompanied by a
jump-like increase in the martensitic transformation temperature
followed by its rapid growth with further deviation from
stoichiometry, while Curie temperature exhibits an ordinary
decrease. These peculiarities of the phase diagram of
Ni$_{2+x}$Mn$_{1-x}$Ga can be explained as follow.

It is well documented in the literature that the compositional
dependence of the martensitic transition temperature $T_m$ in
Ni-Mn-Ga alloys is related to the valence electron concentration
$e/a$, i.e., can be attributed to the Hume-Rothery
mechanism.~\cite{27-c} The martensitic transition takes place when
the contact between the Fermi surface and Brillouin zone boundary
occurs.~\cite{37-w} Such a scenario implies that the change in the
number of valence electrons and the alteration of Brillouine zone
boundary are primary driving forces for the occurrence of
structural instability in these alloys. Neglecting hybridization
effects and other factors such as electronegativity
difference,~\cite{37a-w} one can expect to detect a linear change
of $T_m$ with composition due to a monotonous change in the number
of valence electrons and in the chemical pressure, which has
indeed been observed in limited composition intervals of
Ni$_{2+x}$Mn$_{1-x}$Ga,~\cite{23-v}
Ni$_{2+x}$MnGa$_{1-x}$,~\cite{25-l} and
Ni$_2$Mn$_{1+x}$Ga$_{1-x}$~\cite{24-j} systems. However, this
picture will be no longer valid when approaching Curie point,
because volume magnetostriction considerably affects crystal
lattice parameters. In this sense the peak of the thermal
expansion coefficient observed at Curie point $T_C$ (Fig.~5) can
be regarded as a potential barrier for the increase of $T_m$ which
is "blocked" at the temperature of this peak. Further change of
martensitic transition temperature $T_m$ will correlate with the
change in Curie temperature $T_C$, i.e., $T_m$ and $T_C$ will be
coupled within some compositional interval. To decouple these
phase transitions, it is necessary to reach an $e/a$ value
sufficient for overcoming the barrier caused by the volume
magnetostriction.

Rather a wide compositional interval of Ni$_{2+x}$Mn$_{1-x}$Ga
where $T_m$ and $T_C$ are coupled (Fig.~3) can also be conditioned
by the fact that upon this substitution the conduction electron
density changes more slowly as compared to the substitution of Ni
for Ga or Mn for Ga. Since the increase in the number of valence
electrons is larger in the case of substitution of Ni for Ga, the
compositional interval of coupled magnetostructural phase
transition should be narrower in Ni$_{2+x}$MnGa$_{1-x}$ as
compared to Ni$_{2+x}$Mn$_{1-x}$Ga. It is also worth noting that
the chemical pressure which also can influence the width of the
compositional interval with merged $T_m$ and $T_C$ has different
sign in these two systems (positive in the case of substitution of
Ni for Ga and negative when substituting Ni for Mn).

Coupling of magnetic and martensitic transitions leads to an
unusual behavior of the magnetic transition temperature. Despite
of progressive dilution of the magnetic subsystem occurring upon
substitution of Ni for Mn, the decrease of Curie temperature in $0
\le x \le 0.18$ is followed by its increase in the interval $0.18
\le x \le 0.22$, then $T_C$ becomes independent of composition for
$0.22 \le x \le 0.27$. The expected decrease of $T_C$ is observed
only for $x \ge 0.30$, where $T_m$ is considerably higher then
$T_C$.

The growth of $T_C$ observed in the $0.18 \le x \le 0.22$ interval
is caused by the fact that the exchange interactions in the
martensitic phase is stronger than in the austenitic
phase.~\cite{29-k,38-c} This is supported circumstantially by the
results of $M(H)$ measurements performed at liquid helium
temperature (Fig.~4). These measurements indicated that the
magnetic moment exhibits approximately linear decrease upon
substitution of Ni for Mn; no anomalous behavior is observed in
the interval of compositions $0.18 \le x \le 0.22$. Thus, it can
be suggested that $T_C$ for Ni$_{2.18}$Mn$_{0.82}$Ga $(x = 0.18)$
corresponds to Curie temperature of austenite $T_C^A$, whereas
$T_C$ for Ni$_{2.22}$Mn$_{0.78}$Ga $(x = 0.22)$ corresponds to
Curie temperature of martensite $T_C^M$. Almost constant
temperature of the magnetostructural transition, $T_m \approx T_C
\approx 370$~K, observed in the $0.22 \le x \le 0.27$ alloys, is
probably caused by a competition between increasing electron
population and further dilution of the magnetic subsystem
occurring in the presence of the strong magnetoelastic
interaction.

Assuming that virtual Curie temperature of the austenitic phase
(compositions with $x > 0.18$) decreases linearly in the same
manner as $T_C$ does in the compositional interval $0 \le x \le
0.18$, we can estimate the difference between Curie temperatures
of austenite and martensite. It follows from our data (Fig.~3)
that it can be calculated as a difference between $T_C$ for the $x
= 0.22$ composition and the virtual austenitic Curie temperature
for this composition, $\Delta T_C = T_C^M - T_C^A$ = 370~K --
316~K = 54~K. This value is in good agreement with that estimated
from the comparison of reduced magnetization of austenitic and
martensitic phases.~\cite{29-k} Slightly smaller difference
between $T_C^M$ and $T_C^A$, $\Delta T_C \approx 46$~K, is
obtained when using compositional dependence of the martensitic
Curie temperature estimated from the data for $0.30 \le x \le
0.36$. This can be caused by the fact that for the $0.30 \le x \le
0.36$ compositions the effect of disordering on $T_C^M$ was not
taken into consideration.

In a critical composition $x = 0.30$, $T_m$ and $T_C$ are no
longer coupled, which results in a drastic increase of the
martensitic transformation temperature above 500~K. With further
increase in the Ni excess, $T_m$ rapidly grows and in
Ni$_{2.36}$Mn$_{0.64}$Ga $(x = 0.36)$ the martensitic
transformation occurs at temperatures above 600~K. As evident from
our phase diagram (Fig.~3) and from the data summarized for a
large number of Ni-Mn-Ga compositions,~\cite{27-c} compositional
dependence of $T_m$ differs in the ferromagnetic and the
paramagnetic states. This fact can be related to the difference in
the electronic structure of these phases. Indeed, first principle
calculations~\cite{39-f} have revealed sharp distinctions between
density of states of ferromagnetic and paramagnetic phases.
Specifically, in the ferromagnetic state the Fermi energy passes
through a peak of Ni $d$-states, whereas Fermi level of the
paramagnetic state is located at a high peak of Mn $d$-states.
Although calculations of the electronic structure of paramagnetic
tetragonal phase~\cite{39-f} did not support scenario of a band
Jahn-Teller effect as a driving force for the martensitic phase
transition, it is necessary to stress that the martensitic
transition takes place in paramagnetic state only in the alloys
with a large deviation from stoichiometry, whereas the first
principle calculations were performed for stoichiometric
Ni$_2$MnGa composition.

As compared to the theoretical analysis of phase transitions
presented in Ref.~\onlinecite{23-v}, the theoretical approach
adopted in this work differs in several aspects. It is well known
that anisotropic magnetostriction tends to zero in the vicinity of
Curie point $T_C$; it is also rather small even below $T_C$ as
compared to the volume magnetostriction at temperatures near
$T_C$. Fradkin showed~\cite{34-f} that the volume magnetostriction
has an influence on magnetostructural phase transition only if one
considers a term connecting order parameter $e_1$, responsible for
a volume change at phase transition, with order parameters $e_2$
and $e_3$, responsible for symmetry changes at phase transitions.
Although volume magnetostriction was considered in
Ref.~\onlinecite{23-v} in the Landau functional, the term
connecting $e_1$ and $e_2$, $e_3$ order parameters was not taken
into account. Because of this, the role of the volume
magnetostriction came to a renormalization of the exchange
constant $\alpha$ and, thus, the interaction between magnetic and
structural subsystems was accounted for by the anisotropic
magnetostriction only. As a consequence, the theoretical phase
diagram~\cite{23-v} did not account accurately the width of the
composition interval with coupled magnetostructural phase
transition. Other relevant difference between theoretical analysis
of the present work and that given in Ref.~\onlinecite{23-v}
concerns areas of stability of the phases. Conditions for the
phase equilibrium are different due to the fact that in
Ref.~\onlinecite{23-v} the first cubic anisotropy constant $K$ was
assumed to be negative ($K < 0$), whereas subsequent experimental
results~\cite{40-t} showed that in fact it is positive ($K > 0$).

Present theoretical study, which assumes a linear decrease of
Curie temperature due to the dilution of the magnetic subsystem
and a linear increase of martensitic transition due to the
increase in electron concentration, showed fair agreement with the
experiment for the regions of the phase diagram where $T_m$ and
$T_C$ are decoupled. As for the region characterized by the
coupled magnetostructural phase transition, the applicability of
the theory is limited due to the following reasons. Although
consideration of the volume magnetostriction is shown to be useful
for describing the width of composition interval of
magnetostructural phase transition, just the volume
magnetostriction is responsible for the deviation from a linear
increase of the magnetostructural transition temperature in $0.22
\le x \le 0.27$. It had been suggested in the above discussion
that despite increasing number of valence electrons $T_m$ does not
increase due to the influence of volume magnetostriction. Magnetic
transition, it its turn, will remain coupled to the structural
transition until $T_m$ temperature will not exceed the Curie
temperature of martensite $T_C^M$. Thus, the behavior of
magnetostructural transition temperature in $0.22 \le x \le 0.27$
can be understood as a delicate balance between the change of
Brillouine zone boundary caused by the volume magnetostriction and
an increase in valence electron concentration occurring upon
substitution of Ni for Mn, i.e., by the microscopic arguments
which can not be accounted in the phenomenological approach.

Strictly speaking, the agreement between theoretical and
experimental phase diagram in the interval of compositions $0.18
\le x \le 0.22$ is essentially due to the fact that exchange
interactions are stronger in the martensitic phase than in the
austenitic phase, which leads to the increase of magnetostructural
transition temperature up to $x = 0.22$. In the opposite case of
$T_C^M \le T_C^A$ (experimentally observed in a Ni-Mn-Sn
system~\cite{51-k}) no increase in the magnetostructural
transition temperature should be observed.

\section{Conclusion}

We investigated, experimentally and theoretically, phase diagram
of ferromagnetic shape memory alloys Ni$_{2+x}$Mn$_{1-x}$Ga in the
range of $x$ up to 0.36. Peculiar feature of the phase diagram was
found in a compositional interval $0.18 \le x \le 0.27$. For these
alloys, martensitic and ferromagnetic transitions are merged in a
coupled magnetostructural phase transition from ferromagnetic
austenite to paramagnetic martensite. Due to the difference in
Curie temperatures of austenite and martensite, the temperature of
this phase transition has a non-monotonous dependence on Ni excess
$x$, whereas magnetic moment per formula unit exhibits an ordinary
decrease with the dilution of the Mn magnetic subsystem. The
extended compositional interval of merged $T_m$ and $T_C$ is
suggested to be due to the influence of volume magnetostriction.
The difference in Curie temperatures of martensitic and austenitic
phases estimated from the experimental phase diagram is $\Delta
T_C \sim 50$~K, which is in good agreement with the value obtained
from the comparison of reduced magnetization of austenitic and
martensitic phases.~\cite{29-k}

In the alloys with $x \ge 0.30$, $T_m$ and $T_C$ are no longer
coupled, and the martensitic transformation takes place at
temperatures above 500~K. Since such materials are important for
high-temperature shape memory alloys, it can be suggested that in
the $0.30 \le x \le 0.36$ alloys further increase in the
martensitic transformation temperature can be attained by the
substitution of Ga for Ni or for Mn.

\section*{Acknowledgments}

This work was partially supported by Russian Foundation for Basic
Research (Grants Nos. 03-02-17443, 04-02-81058, and 03-02-39006),
by the RFBR--JSPS joint project No. 05-02-19935 and by RF and
CRDF, post-doctoral grant Y2-P-05-19.

\end{document}